\documentclass[11pt,a4paper]{article}
\pdfoutput=1
\usepackage{jheppub}
\usepackage{amsmath}%
\usepackage{amsfonts}%
\usepackage{amssymb}%
\usepackage{graphicx}
\usepackage{tikz-feynman}
\usepackage{subfig}
\usepackage{slashed}
\usepackage{comment}

\newcommand{\be}{\begin{equation}}
\newcommand{\ee}{\end{equation}}

\title{Disformally Coupled Scalar Fields and  Inspiralling Trajectories}

\author[a]{Philippe Brax}
 \author[b]{Anne-Christine Davis}
 \author[c]{Adrien Kuntz}

\affiliation[a]{Institut de Physique Th\'eorique, Universit\'e Paris-Saclay, CEA,CNRS,\\
F-91191Gif sur Yvette, France }
\affiliation[b]{DAMTP, Centre for Mathematical Sciences, University of Cambridge,
  CB3 0WA, UK}
 \affiliation[c]{Centre de Physique Th\'{e}orique, Aix-Marseille Universit\'{e},  Universit\'{e} de Toulon, CNRS UMR 7332, 13288 Marseille, France}

\emailAdd{philippe.brax@cea.fr}
\emailAdd{a.c.davis@damtp.cam.ac.uk}
\emailAdd{adrien.kuntz@univ-amu.fr}

\abstract{We show how a nearly massless scalar field conformally and disformally coupled to matter can affect the dynamics of two bodies in their inspiralling phase before merging. We discuss both the conservative dynamics, e.g. how the energy of the bound system is corrected by the conformal and disformal interactions, and the dissipative part where scalars and gravitons are emitted. The first disformal correction to the Einstein-Infeld-Hoffmann Lagrangian is obtained using both the Fokker method relying on the equations of motion and an Effective Field Theory approach using Feynman diagrams. This leads to a correction to the energy functional at the 2PN level for eccentric orbits,   which vanishes for circular orbits up to the 7PN order. The dissipative power from the disformal interaction gives a correction to the monopole and quadrupole terms in the presence of a conformal coupling. Although this correction vanishes for circular orbits at leading order, this is not the case for elliptical orbits allowing us to derive a bound on the disformal coupling from the time drift of the period for the Hulse-Taylor binary pulsar, which is slightly stronger than the one from fifth force tests. We conclude that the prospect of observing disformal effects for inspiralling systems lies in the accurate monitoring of eccentric trajectories as expected for the LISA experiment.  }

\begin{document}

\maketitle
\flushbottom

\section{Introduction}
Scalar fields \cite{Copeland:2006wr} are ubiquitous in cosmology and have been intensely studied as plausible candidates for inflation \cite{Linde:1983gd}, dark matter and dark energy \cite{Khoury:2013tda}. They have an indirect
influence on the dynamics of the Universe and their plausible presence could be inferred from cosmological data, past and future \cite{Amendola:2012ys}. On the other hand, one would like to
test the existence of scalar fields in the Universe in a direct manner using experimental techniques involving matter as convenient probes. This can only be envisaged when the putative scalars couple to ordinary matter \cite{Brax:2013ida}.

One can distinguish two types of couplings to matter. The first class involves conformal couplings and lead to direct interactions such as Yukawa contact terms \cite{Damour:1992we} or quadratic terms in the symmetron model \cite{Hinterbichler:2010es}. Another class falling within the Bekenstein classification \cite{Bekenstein:1992pj}  is of the derivative nature and involves a pair of scalar derivatives coupled to the energy-momentum tensor of matter. This type of coupling is known as a disformal term \cite{Brax:2014d} and appears naturally in constructions of scalar models of the Horndeski form \cite{Horndeski:1974wa,Langlois:2015cwa}. Disformal couplings have been investigated in a variety of contexts \cite{Kaloper:2003yf, Koivisto:2008ak, Zumalacarregui:2010wj, Koivisto:2012za, vandeBruck:2013yxa, Brax:2013nsa, Brax:2013nsa, Sakstein:2014aca, Sakstein:2014isa, Sakstein:2015jca, Brax:2015hma, vandeBruck:2015ida} and bounded quite strongly.

In this paper, we calculate the first disformal correction to the dynamics of a two body bound system \cite{Brax:2018bow} such as neutron stars or black holes. This correction has two origins. The first one follows from the scalar interaction between the bodies and results in a modification of the Einstein-Infeld-Hoffmann Lagrangian. The second one comes from the power emitted by the two body system into scalars; the presence of the disformal interaction opens up new channels for this dissipative effects leading to a correction to the Peter-Mathews formula \cite{PhysRev.131.435}. We establish the remarkable property that for circular orbits the effects of the disformal coupling vanish up to the seven Post-Newtonian (7PN) order for the conservative part of the dynamics, and 6PN for the dissipative dynamics. Hence only eccentric orbits are sensitive to the presence of the disformal interaction. This can be intuitively understood from the fact that the disformal coupling to a point-like object is a time derivative, i.e it reads
\begin{equation}
\int d\tau \left( \frac{d\phi}{d\tau} \right)^2
\end{equation}
where $\tau$ is the proper time of the object we consider. As this time derivative vanishes when the  relative distance  $r$ between two orbiting objects is constant for circular orbits, we conclude that the effects of the disformal coupling will only appear when relativistic effects must be taken into account.

Finally we apply our results  to  the rate of change of the period for the Hulse-Taylor binary pulsar \cite{Hulse:1974eb}. We deduce a bound on the disformal coupling which is slighly tighter than the one from tests of new gravitational forces in the sub-millimetric range by the E\"ot-wash experiment \cite{Adelberger:2002ic, Brax:2014vva}.  These bounds on the disformal coupling imply that the corrections for elliptical orbits will be small in the inspiralling phase prior to the merging of two objects such as neutron stars or black holes. Only when the two bodies are close enough does the disformal interaction become large. This is a regime which is beyond the scope of this article and which should deserve further investigation.

This paper is arranged as follows. In Sec. \ref{sec:conservative} we compute the first disformal correction to the conservative two-body dynamics using two different approaches : the Fokker effective action and the Effective Field Theory approach. We examine in more detail the case of a circular orbit, for which the disformal correction vanishes up to high PN order. Section \ref{sec:disf} then presents the radiative part of the dynamics. We compute the radiated power into scalars and derive a bound on the disformal energy from the Hulse-Taylor pulsar data. We conclude in Section \ref{sec:conclusion}. Appendix \ref{app:A} contains details concerning the Feynman rules and the calculation of the diagrams in the main text, and App. \ref{app:B} contains details about the radiation from point particles sources. Our metric convention is $(-+++)$ and we use Planck's mass normalization $m_\mathrm{Pl}^2 = 1/(8\pi G_N)$.

%In a first paper we study the conservative dynamics from two points of view: first using Fokker's method and the equations of motions, then using Feynman diagrams. Then we focus on the dissipative dynamics. We then conclude. Appendices contain details about diagrammatics and the radiation from point particles.

\section{Disformally coupled scalars and two body conservative dynamics} \label{sec:conservative}
\subsection{The model} \label{sec:model}

We are interested in the dynamics of macroscopic bodies interacting via both gravity and a massless scalar field.
The dynamics are specified by
 the total action which reads
\be
S=\int d^4x \sqrt{-g_E}\left ( \frac{R_E}{16\pi G_N}-\frac{1}{2} (\partial\phi)^2\right ) +S_m(\psi_i, g_{\mu\nu})
\ee
in the Einstein frame for the Einstein-Hilbert action and a massless scalar field. The matter fields are denoted by $\psi_i$ and their action is $S_m$. In the following we will take the matter action to be the one of point-like particles. This will give an appropriate description of the dynamics of macroscopic objects like neutron stars as long as
finite size effects can be neglected.

The coupling between a massless scalar field and matter can be mediated by a metric which differs from the one appearing in the Einstein-Hilbert action. In the following we will focus
on metrics of the Bekenstein form where both conformal and disformal couplings are present. In general this could involve two functions $A(\phi,X)$ and $B(\phi,X)$ of the scalar field $\phi$ and its derivatives $X=-\frac{1}{2} (\partial \phi)^2$ where the contraction is performed with the Einstein metric $g_{\mu\nu}^E$. Here
we consider the scalar interaction between  moving bodies when the coupling between matter and the scalar field is mediated by the metric
\be
g_{\mu\nu}= A^2(\phi) g_{\mu\nu}^E+ B^2(\phi) \partial_\mu\phi \partial_\nu \phi
\label{eq:jordan_frame_metric}
\ee
where we assume that both the conformal and disformal factors $A(\phi)$ and $B(\phi)$ are independent of $X$. More precisely we will focus on the simplest type of conformal coupling
\be
A(\phi)= e^{\beta \phi/m_{\rm Pl}}
\ee
leading to a Yukawa interaction of coupling strength $\beta$ with matter.  The coupling $\beta$ is strongly constrained for a massless scalar, as no screening mechanism is involved, by the Cassini bound $\beta^2 \lesssim 10^{-5}$ \cite{Bertotti:2003rm}. In the general case we  assume that the conformal coupling $A$ can be expanded in $\phi/m_\mathrm{Pl}$. As the scalar field generated by a massive body of mass $m$ can be  approximated by
\begin{equation}
\phi \simeq \frac{\beta m}{4\pi m_\mathrm{Pl} r}
\end{equation}
where $r$ is the distance to the source,
the condition $\phi/m_\mathrm{Pl} \lesssim 1$  translates into $r \gtrsim \tilde r_s$ where
\begin{equation}
\tilde r_s = \beta \frac{m}{4\pi m_\mathrm{Pl}^2}
\end{equation}
is the Schwarzschild radius of the source $r_s = 2 G_N m$  corrected by a factor $\beta$. As our perturbative treatment is only valid well outside the Schwarzschild radius
of the moving bodies and as $\beta \ll 1$ from the Cassini bound, we conclude that the series expansion in $\phi/m_\mathrm{Pl}$ is always valid where the
perturbative treatment of the motion of moving objects can be applied.
We will see shortly that the disformal coupling introduces  a new nonlinear scale $r_\star$, which we can tune to be of the same order as the Schwarzschild radius for objects of the mass of the sun, leading to interesting departures from GR that should appear in the waveform generated by two inspiralling black holes or neutron stars.

We consider the disformal term to be a power series in $\phi/m_\mathrm{Pl}$. At leading order we have
\be
B^2(\phi) = \frac{2}{ m^2_{\rm Pl} \Lambda^2}
\ee
where the new scale $\Lambda$ characterises the strength of the disformal interaction. Close to a non-relativistic source of mass $m$, $A(\phi) \sim g_{\mu \nu}^E \sim 1$. By requiring that the disformal factor in eq. \eqref{eq:jordan_frame_metric} should be less that the conformal one in order to recover Newtonian mechanics, we obtain the following bound
\begin{equation} \label{eq:small_ratio}
\frac{\beta m}{4\pi m_\mathrm{Pl}^2 \Lambda r^2} \lesssim 1
\end{equation}
which simply means that our theory is valid in the range $r \gtrsim r_\star$, where $r_\star$ is the nonlinear radius
\begin{equation} \label{eq:nonlinear_radius}
r_\star = \sqrt{\frac{\beta m}{4 \pi m_\mathrm{Pl}^2 \Lambda}} \; .
\end{equation}
For distances $r < r_*$, the perturbative calculation that we use in this article will be invalid. One can tune the mass scale $\Lambda$ so that $r_*$ is of the same order as the Schwarzschild radius for an object like the Sun which gives then the scaling relation
$
r_\star = r_{s, \odot} \sqrt{\frac{m}{m_\odot}}
$
for objects of different masses.
One may worry that this could  give   too large a nonlinear radius for small mass objects, however this is not the case. For example, a body of mass $m=1$ kg would have a nonlinear radius  $r_* = 10^{-12}$ m, consequently usual matter has a nonlinear radius well within its extension.
The lower bound on $\Lambda$ obtained by requiring $r_* \lesssim  r_{s, \odot}$ is
\begin{equation}
\Lambda \gtrsim 10^{-13} \; \mathrm{eV}
\end{equation}
where we have taken $\beta^2 \simeq 10^{-5}$ from the Cassini bound.
This should  be compared with the bound obtained by, e.g.,  torsion balance experiments $\Lambda \gtrsim 10^{-18}$ eV \cite{Brax:2014vva}. If one takes also into account the constraints coming from particle physics experiments, then the value of $\Lambda$ should be much higher, $\Lambda \gtrsim 10^{-4}$~eV \cite{Brax:2014vva}, leading to a non-linear radius well within the Schwartzschild radius of solar mass objects. We chose nonetheless to carry our analysis independently of collider constraints.

Let us now turn to the equations of motion of the theory. The gravitational dynamics are dictated by the Einstein equation
\be
R_{\mu\nu}-\frac{1}{2} R g_{\mu\nu}= 8\pi G_N (T_{\mu\nu}+ T^\phi_{\mu\nu})
\ee
where the matter energy-momentum tensor in the Einstein frame  is
\be
T_{\mu\nu}=-\frac{2}{\sqrt{-g^E}} \frac{\delta S_m}{\delta g^{\mu\nu}_E}.
\ee
The corresponding one for the scalar field is given by
\be
T^\phi_{\mu\nu}= \partial_\mu\phi \partial_\nu \phi-\frac{(\partial\phi)^2}{2} g_{\mu\nu}^E.
\ee
The dynamics of the scalar field are given by the Klein-Gordon equation
\be
\Box \phi= -\beta\frac{T}{m_{\rm Pl}} +\frac{2}{\Lambda^2 m_\mathrm{Pl}^2} \partial_\mu (  \partial_\nu\phi T^{\mu\nu})
\ee
where we consider the metric to be flat at leading order. In the following we will analyse the solutions of the Klein-Gordon equation when matter is given by point-like sources.

\subsection{Scalar field generated by two moving bodies}

For point sources of mass $m_\alpha$ the energy momentum tensors read
\be
T^{\mu\nu}_\alpha = m_\alpha \int d\tau A(\phi) u_\alpha^\mu u_\alpha^\nu \delta^{(4)} (x^\mu -x_\alpha^\mu(\tau))
\ee
where $\tau$ is the proper time of each particle in the Einstein frame such that
$
u^\mu=\frac{dx_\alpha^\mu}{d\tau}
$
and $u^\mu_\alpha u_\mu^\alpha=-1$. Notice that, as we work in the Einstein frame, the masses of the particles become $m_\alpha A(\phi)$ which is field dependent.
In this Section we work at leading order in $\beta$ and $1/\Lambda^2$ which implies that we can safely take $A(\phi)\sim 1$ in the Klein-Gordon equation. Higher order contributions can be taken into account in the diagrammatic approach we will use in Sec. \ref{sec:diagrammatic_approach}.
When two moving bodies are present, the solution to the Klein-Gordon equation can be obtained in two steps \cite{Brax:2018bow}.
The first step consists in solving the Klein-Gordon equation with no disformal coupling
\be
\Box \phi^{(0)}= -\beta\frac{T^A +T^B}{m_{\rm Pl}}
\ee
where the energy momentum tensor contains both the parts from particles $A$ and $B$.
The solution is simply given by the linear combination
\be
\phi^{(0)}= \phi^{(0)}_A+ \phi^{(0)}_B
\ee
where we have
\be
\phi^{(0)}_{A,B}( x)= -\frac{\beta m_{A,B}(1-\frac{\vec v_{A,B}^2}{2}+\frac{\vec v_{A,B\perp}^2}{2})}{4\pi m_{\rm Pl}\vert \vec x-\vec x_{A,B}\vert}.
\label{corB}
\ee
and we have defined $\vec v_{A\perp}= \vec v_A - \vec n_A (\vec v_A . \vec n_A)$ where $\vec n_A = (\vec x - \vec x_A)/\vert \vec x -\vec x_A\vert$. We have explicitly worked at lowest order in the velocities.
This solution sources the next step in the iteration process
\be
\Box \delta\phi^{(0)}= \frac{2}{m_{\rm Pl}^2 \Lambda^2} \partial_\mu (  \partial_\nu\phi^{(0)} T^{\mu\nu})
\ee
and leads to four contributions
\be
\delta \phi^{(0)}_{\alpha\beta}(x)= -\frac{m_\alpha}{2\pi m_{\rm Pl}^2 \Lambda^2 \gamma_\alpha }\frac{\partial_\mu (\partial_\nu \phi_\beta^{(0)}(x_\alpha) u_\alpha^\mu u_\alpha^\nu)}{ \vert \vec x-\vec x_\alpha\vert}
\ee
where $\alpha,\beta=A,B$ and $\gamma_\alpha= \frac{1}{\sqrt{1-\vec v_\alpha^2}}$.
It turns out then that $\delta \phi^{(0)}_{AA}$ and $\delta \phi^{(0)}_{BB}$ both vanish whilst
\begin{eqnarray}
&&\delta \phi^{(0)}_{AB}=-\frac{\beta G_N m_A m_B}{\pi m_{\rm Pl}} \frac{(\vec a_A -\vec a_B). \vec n_{AB}+(\vec v_A -\vec v_B)^2 -3 (\vec n_{AB}.(\vec v_A-\vec v_B))^2}{\Lambda^2\vert \vec x-\vec x_A\vert \vert \vec x_B-\vec x_A\vert^3}\nonumber \\
&&\delta \phi^{(0)}_{BA}=-\frac{\beta G_N m_A m_B}{\pi m_{\rm Pl}} \frac{(\vec a_A -\vec a_B). \vec n_{AB}+(\vec v_A -\vec v_B)^2 -3 (\vec n_{AB}.(\vec v_A-\vec v_B))^2}{\Lambda^2\vert \vec x-\vec x_B\vert \vert \vec x_B-\vec x_A\vert^3}\nonumber \\
\label{corA}
\end{eqnarray}
where $\vec n_{AB}$ is the unit vector between $A$ and $B$ . Notice that the acceleration appears at this order where $\vec a_\alpha= \frac{d\vec v_\alpha}{dt}$. Higher order contributions are obtained by solving iteratively
\be
\Box \delta\phi^{(n+1)}= \frac{2}{\Lambda^2 m_{\rm Pl}^2} \partial_\mu   (\partial_\nu\delta \phi^{(n)} T^{\mu\nu})
\ee
and summing
\be
\phi=\phi^{(0)} + \sum _{n\ge 0} \delta \phi^{(n)}.
\ee
In the following we shall focus on $n=0$ only.

\subsection{The Fokker effective action}

We can now calculate the effective action for the motion of two bodies. This is obtained by evaluating the action of the scalar field together with the gravitational action when the scalar field and the metric satisfy their equations of motion. In this Section we first consider the scalar part of the action. At leading order the matter Lagrangian can be expanded and corresponds to
\be
S_m= \int d^4 x \left ( \frac{\beta}{m_{\rm Pl}}(\phi_A+\phi_B) (T_A +T_B) +\frac{1}{m_{\rm Pl}^2 \Lambda^2} \partial_\mu (\phi_A+\phi_B) \partial_\nu (\phi_A +\phi_B)( T^{\mu\nu}_A +T^{\mu\nu}_B)\right )
\ee
Similarly the scalar action can be obtained after integration by parts and using the Klein-Gordon equation
\begin{align}
\begin{split}
S_{\rm scalar} &= -\frac{1}{2}\int d^4 x \left ( +\frac{\beta}{m_{\rm Pl}}(\phi_A+\phi_B) (T_A +T_B) \right. \\
&+ \left. \frac{2}{m_{\rm Pl}^2 \Lambda^2} \partial_\mu (\phi_A+\phi_B) \partial_\nu (\phi_A +\phi_B)( T^{\mu\nu}_A +T^{\mu\nu}_B)\right )
\end{split}
\end{align}
so that the part of the effective action which comes from the scalar field reads
\be
S_{\rm scalar}= \frac{1}{2}\int d^4 x  \frac{\beta}{m_{\rm Pl}}(\phi_A+\phi_B) (T_A +T_B).
\label{scalar}
\ee
Of course, there are self-energy divergences in this expression which must be removed. For instance the scalar field in this action around $\vec x_A$ is
\be
\phi(x)= \bar \phi_B(x) + {\cal}{O}(\frac{1}{\vert \vec x-\vec x_A\vert}).
\ee
where explicitly
\be
\bar\phi_B (x)= \phi^{(0)}_B(x) +\delta \phi^{(0)}_{BA}(x)
\ee
is the field generated by the particle $B$ evaluated around the  particle $A$.
The scalar action (\ref{scalar})  can be explicitly evaluated and gives the scalar part of the Lagrangian
\begin{eqnarray}
&&{\cal L}_{S}=-\frac{\beta^2 G_N m_A m_B}{\vert x_B-x_A\vert}(\vec v_A^2 +\vec v_B^2)  +\frac{2\beta^2 G_N m_A m_B}{\vert x_B-x_A\vert}(1+\frac{\vec v_{A\perp}.\vec v_{B.\perp}}{2})\nonumber \\
&&+{4\beta^2 G^2_N}m_Am_B (m_A+m_B)  \frac{((\vec a_A -\vec a_B). \vec n_{AB}+ (\vec v_A-\vec v_B))^2- 3((\vec v_A-\vec v_B).\vec n_{AB})^2}{\Lambda^2\vert x_A -x_B\vert^4}\nonumber \\
\end{eqnarray}
which contributes to the effective dynamics of the two body system. The remaining part of the effective action can be obtained by evaluating the Einstein-Hilbert action for the gravitational fields sourced by two bodies \cite{Damour:1992we}. This yields
\begin{eqnarray}
&& {\cal L}_{AB}=
\frac{1}{2} m_A \vec v_A^2 +\frac{1}{2} m_B \vec v_B^2  -m_A-m_B  +\frac{m_A}{8}  \vec v_A^4 + \frac{m_B}{8} \vec v_B^4+ \frac{G_N(1+2\beta^2) m_Am_B}{\vert \vec x_B-\vec x_A\vert}\nonumber \\
&& + \frac{G_N m_A m_B}{2\vert \vec x_A -\vec x_B\vert}((3-2\beta^2) \vec v_A^2 +(3-2\beta^2) \vec v_B^2 -8 \vec v_A.\vec v_B + (1+2\beta^2) \vec v_{A\perp}.\vec v_{B\perp})\nonumber\\
 &&+ {4\beta^2 G^2_N}m_Am_B (m_A+m_B)  \frac{((\vec a_A -\vec a_B). \vec n_{AB}+(\vec v_A-\vec v_B))^2- 3((\vec v_A-\vec v_B).\vec n_{AB})^2}{\Lambda^2\vert \vec x_A -\vec x_B\vert^4}\nonumber\\
&& -\frac{G_N^2(1+2\beta^2)^2 m_Am_B (m_A+m_B)}{2 r^2} \nonumber \\
\label{uii}
\end{eqnarray}
where $\vec v_{\alpha \perp}= \vec v_\alpha  - (\vec v_\alpha. \vec n_{AB}) \vec n_{AB}$, and $\vec v_{A\perp}. \vec v_{B\perp}= \vec v_A. \vec v_B- (\vec v_A. \vec n_{AB}) (\vec v_B. \vec n_{AB})$.
which has been truncated at leading order in $G_N/\Lambda^2$. This is the  Einstein-Hoffman-Infeld Lagrangian including the effects of both a conformal and a disformal coupling.
In fact this action can be reformulated thanks to the identity
\be
\frac{d}{dt}( \frac{\vec n_{AB}.(\vec v_A -\vec v_B)}{\vert \vec x_A -\vec x_B\vert^3}) +\frac{(\vec n_{AB}.(\vec v_A -\vec v_B))^2}{\vert \vec x_A -\vec x_B\vert^4}=
\frac{((\vec a_A -\vec a_B). \vec n_{AB}+(\vec v_A-\vec v_B))^2- 3((\vec v_A-\vec v_B).\vec n_{AB})^2}{\vert \vec x_A -\vec x_B\vert^4}
\ee
implying that up to a total time derivative which plays no role in the dynamics of the two bodies, the effective Lagrangian becomes
\begin{eqnarray}
&& {\cal L}_{AB}=
\frac{1}{2} m_A \vec v_A^2 +\frac{1}{2} m_B \vec v_B^2  -m_A-m_B  +\frac{m_A}{8}  \vec v_A^4 + \frac{m_B}{8} \vec v_B^8+ \frac{G_N(1+2\beta^2) m_Am_B}{\vert \vec x_B-\vec x_A\vert}\nonumber \\
&& + \frac{G_N m_A m_B}{2\vert \vec x_A -\vec x_B\vert}((3-2\beta^2) \vec v_A^2 +(3-2\beta^2) \vec v_B^2 -8 \vec v_A.\vec v_B + (1+2\beta^2) \vec v_{A\perp}.\vec v_{B\perp})\nonumber\\
 &&+ {4\beta^2 G^2_N}m_Am_B (m_A+m_B) \frac{(\vec n_{AB}.(\vec v_A -\vec v_B))^2}{\Lambda^2 \vert \vec x_A -\vec x_B\vert^4}\nonumber\\
&& -\frac{G_N^2(1+2\beta^2)^2 m_Am_B (m_A+m_B)}{2 r^2} \nonumber \\
\label{uii}
\end{eqnarray}
with only one term coming from the disformal coupling and correcting the Einstein-Infeld-Hoffmann Lagrangian.

\subsection{Equivalence with the diagrammatic approach} \label{sec:diagrammatic_approach}
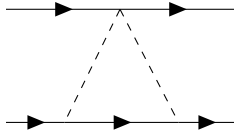
\begin{figure}
	\centering
		\begin{tikzpicture}
			\begin{feynman}
				\vertex (i1);
				\vertex [right=of i1] (a);
				\vertex [right=of a] (f1);
				\vertex [below=of i1] (i2);
				\vertex [right=2em of i2] (b);
				\vertex [right=of b] (c);
				\vertex [below=of f1] (f2);
				
				\diagram*{
				(i1) -- [fermion] (a) -- [fermion] (f1),
				(i2) -- [fermion] (b) -- [fermion] (c) -- [fermion] (f2),
				(a) -- [scalar] {(b), (c)},
				};
				
			\end{feynman}
		
		\end{tikzpicture}
\caption{Feynman diagram corresponding to the first disformal correction to the conservative dynamics (it should also be included with its symmetric counterpart). The upper vertex is the disformal one in eq. \eqref{eq:disformal_vertex}.}
\label{fig:disformal}
\end{figure}

The previous result obtained using the Fokker method, i.e. calculating the field and replacing  it into the action,  can be recovered using a field theory approach along the lines of Non-Relativistic General Relativity \cite{goldberger_effective_2006, porto_effective_2016}. A generalisation of this formalism including scalar fields can be found in  \cite{Kuntz:2019zef}. Here we use the fact that the action of a point-particle $\alpha$ can be written as
\begin{equation}
S_{m, \alpha} = -m_\alpha \int d\tau_\alpha
\end{equation}
where $d\tau_\alpha^2 = - g_{\mu \nu} dx_\alpha^\mu dx_\alpha^\nu$, and $g_{\mu \nu}$ is the Jordan frame metric, given in eq. \eqref{eq:jordan_frame_metric}. Expanding the conformal factor for weak field values in $\phi/m_\mathrm{Pl}$, the Jordan frame proper time $d\tau_\alpha$ becomes  related to the Einstein proper time $d\tau^E_\alpha$ \textit{via}
\begin{equation}
d\tau_\alpha = d\tau^E_\alpha \left(1+\frac{\beta \phi}{m_\mathrm{Pl}} - \frac{1}{\Lambda^2 m_\mathrm{Pl}^2} \left( \partial_\mu \phi u^\mu_\alpha \right)^2 \right)
\end{equation}
where $u^\mu_\alpha$ is the four-velocity of the point-particle. We have expanded the conformal factor $A$ only to first order in $\phi$, as the higher-order contributions will not be relevant. By using this expression in the action, we find that the interaction vertices are
\begin{equation}
S_{m, \alpha} \supset - \frac{\beta m_\alpha}{m_\mathrm{Pl}} \int d\tau_\alpha \phi + \frac{m_\alpha}{\Lambda^2 m_\mathrm{Pl}^2} \int d\tau_\alpha \left( \partial_\mu \phi u^\mu_\alpha \right)^2 \; .
\label{eq:disformal_vertex}
\end{equation}
We can now use these two vertices to build Feynman diagrams. We refer the reader to \cite{Kuntz:2019zef} for a thorough discussion of the methodology used in this field theory approach to the problem of motion in scalar-tensor theories. A quick summary of the Feynman rules is given in App. \ref{app:A}. The lowest-order contribution of the disformal vertex to the two-body Lagrangian is given by the diagram of Figure \ref{fig:disformal}, which can be calculated to be
\begin{equation}
\mathrm{Fig} \; \ref{fig:disformal} = \frac{4 \beta^2 G_N^2 m_A m_B (m_A + m_B)}{\Lambda^2} \frac{(\vec n_{AB} \cdot (\vec v_A - \vec v_B))^2}{|\vec x_A - \vec x_B|^4}.
\label{eq:disformal_energy}
\end{equation}
See the App. \ref{app:A} for the explicit calculation. This is exactly the same interaction term as the one calculated using the Fokker method and the equations of motion.

\subsection{Energy for elliptic orbits} \label{sec:planar_dynamics}

In Newtonian mechanics, the two body problem is easily solved and trajectories in the centre of mass frame become elliptical for bound systems with planar trajectories parameterised as
\begin{equation} \label{eq:elliptic_trajectory}
r(\psi) = \frac{a(1-e^2)}{1+e \cos \psi}
\end{equation}
where $a$ is the semi-major axis, $e$ the eccentricity, and $\psi$ is the angle of the trajectory in the plane defined by the motion. We also define $p=a(1-e^2)$.  The conservation of angular momentum leads to the following expression for the time derivative of $\psi$
\begin{equation} \label{eq:psidot}
\dot \psi = \left( \frac{\tilde G M}{p^3} \right)^{1/2} (1+e \cos \psi)^{2}
\end{equation}
where $\tilde G = G_N(1+2\beta^2)$ is the renormalised Newton constant that appears in the gravitational force when the leading effect of the conformal interaction is taken into account, i.e. the strength of the Newtonian force is enhanced compared to the purely gravitational case.
Defining by $M=m_A+m_B$ the total mass of the two bodies and $\mu= \frac{m_A m_B}{m_A+m_B}$ the reduced mass, the energy of the system becomes
\be
E= -\frac{\tilde G M\mu}{2a} \; .
\ee

The disformal interaction produces a correction to this Newtonian energy. From the expression of the disformal Lagrangian \eqref{eq:disformal_energy}, this correction to the energy is
\be
\delta E= -\frac{4\beta^2 G_N^2 \mu M^2}{\Lambda^2} \frac{\dot r^2}{r^4} \; .
\ee
where $\dot r = \frac{d r}{dt}$. For elliptical trajectories \eqref{eq:elliptic_trajectory}, this becomes
\be
\delta E= -\frac{4\beta^2 }{(1+2\beta^2)^2} \frac{\mu \tilde G^3 M^3}{\Lambda^2 p^5} e^2 \sin ^2 \psi (1+e \cos \psi)^4
\ee
Over one period $T$ of the system, the mean energy correction is defined as
\begin{align}
\begin{split}
\left \langle \delta E \right \rangle &= \frac{1}{T} \int_0^T dt \; \delta E(t) \\
 &= \frac{1}{2\pi} (1-e^2)^{3/2} \int_0^{2\pi} d\psi \frac{\delta E(\psi)}{(1+e\cos \psi)^{2}}
\end{split}
\end{align}
where we have used the time derivative of $\psi$ in eq. \eqref{eq:psidot} and $T= 2\pi \sqrt{\frac{ a^3}{\tilde G M}}$  using Kepler's third law. This gives for the correction to the energy due to the disformal interaction
\begin{equation} \label{eq:disformal_energy_elliptic}
\frac{\left \langle \delta E \right \rangle}{E} = \frac{4\beta^2}{(1+2\beta^2)^2} \frac{\tilde G^2 M^2}{\Lambda^2 a^4} \frac{e^2(1+\frac{e^2}{4})}{(1-e^2)^{7/2}} \; .
\end{equation}
A few comments are in order. First, note that the disformal energy \eqref{eq:disformal_energy_elliptic} vanishes for circular orbits where $e=0$. Unfortunately, circular trajectories are almost always expected for the LIGO/Virgo detector, as the emission of GW circularises the trajectory \cite{PhysRev.131.435}. However, the space interferometer LISA should be sensitive to trajectories for Extreme Mass Ratio Inspirals, which are expected to be elliptical \cite{AmaroSeoane:2007aw} and so could provide a constraint on the disformal energy. Second, note that the disformal energy is proportional to the conformal factor $\beta^2$ which is strongly constrained by the Cassini bound as $\beta^2 \lesssim 10^{-5}$. Lastly, we can rewrite the dimensionless parameter that appears in the ratio of energies in two different suggestive ways ; the first one is
\begin{equation}
\frac{\beta^2 \tilde G^2 M^2}{\Lambda^2 a^4} \sim \frac{\beta^2 v^4}{\Lambda^2 r^2}
\end{equation}
which simply follows from the virial theorem. This shows that the disformal term induces a 2PN correction to the energy. As the best bound on $\Lambda$ that we will get from the Hulse-Taylor binary pulsar energy loss is $\Lambda \gtrsim 10^{-17}$ eV, the correction to the 2PN Hamiltonian of GR shows up for distances
\begin{equation}
r \sim \frac{\beta }{\Lambda}  \lesssim 1000 \; \mathrm{km} \;
\end{equation}
We stress that this is not in tension with the LIGO/Virgo observation of GW, as the observed orbits are circular and the disformal energy vanishes in this case (we will be more precise on this statement in Sec. \ref{sec:circular}).

A second suggestive way to rewrite the dimensionless parameter is
\begin{equation}
\frac{\beta^2 \tilde G^2 M^2}{\Lambda^2 a^4} = \left( \frac{r_\star}{a} \right)^4
\end{equation}
where $r_*$ is the nonlinear radius \eqref{eq:nonlinear_radius} associated with the total mass $M$, introduced in Sec. \ref{sec:model}. Hence at the limit of validity of the perturbative treatment when $a\simeq r_\star$, this ratio could become close to unity. When the non-linear radius is tuned to be of the order of the Schwarzschild radius for solar mass objects, and close to the merger when the two objects  approach their Schwarzschild radius in relative distance  this parameter will become close to unity and nonperturbative effects should appear. A more detailed analysis of this regime is certainly worth although beyond the scope of the present work.

%where we have introduces the parameter
%\be
%y= \frac{\beta^2 G_N M}{a^3 \Lambda^2}
%\ee
%which must be less than one for the disformal corrections to be smaller than the Newtonian effects and $\Phi= \frac{G_N M}{p}$ the typical  Newtonian potential along the trajectory. As $y\lesssim 1$ and $\Phi \lesssim 1$ for inspiralling objects, the correction to the energy due to the disformal coupling is typically bounded by $e^2$. For nearly circular orbits this vanishes and no correction to the energy appears. For orbits which are elliptical, the order of magnitude is governed by both the Newtonian potential and the expansion parameter $y$. For objects such a neutron stars with a typical mass $M\simeq 2 {\rm M}_\odot$, the Newtonian potential for $r\simeq 1000$ km is $\Phi\sim 10^{-2}$ and taking $\beta^2= 10^{-5}$ to comply with solar system tests and $\Lambda \gtrsim 10^{-11}$ eV, we find that for similar distances $ y\lesssim 10^{-12}$.
%It is only when our perturbative treatment breaks down with much smaller distances that both $\Phi$ and $y$ increasing one could surmise that the disformal effects could become relevant. A more detailed analysis of this regime is beyond the present work.

\subsection{Circular orbits} \label{sec:circular}

The first disformal contribution to the relativistic Lagrangian in the case of a circular orbit vanishes. In this part we will show that this vanishing is also valid at higher order in the Post-Newtonian (PN) expansion. On the other hand radiation reaction at  2.5PN order renders the two-body orbit inspiralling instead of circular, which will eventually gives rise to  a non-zero contribution of the  disformal coupling to the equations of motion. In this Section we investigate the correction to the two-body Lagrangian which do  not vanish for circular orbits.

Let us first notice that the disformal vertex can be rewritten as
\begin{equation}
\frac{m_\alpha}{\Lambda^2 m_\mathrm{Pl}^2} \int d\tau_\alpha \left( \frac{d\phi(x_\alpha)}{d\tau_\alpha} \right)^2
\end{equation}
where the total derivative of $\phi$ is taken along the path of the particle $x_\alpha(\tau_\alpha)$. Intuitively, the fact that the disformal coupling does not contribute for circular orbits can be directly seen from this vertex. Indeed, since the scalar field depends on the relative distance $r = |\vec x_A - \vec x_B|$ which is constant for a circular orbit, the vertex will vanish because of the presence of a time derivative. Let us now be more precise about this fact.

 Consider a general Feynman diagram involving a disformal vertex associated to the first particle and any number of other vertices (including possibly disformal ones), as shown in Figure \ref{fig:disformal_general}. Ignoring the numerical factors, the amplitude of this diagram  can be written as
\begin{figure}
	\centering
		\begin{tikzpicture}
			\begin{feynman}
				\vertex (i1);
				\vertex [right=of i1] (a);
				\vertex [right=of a] (f1);
				\vertex [below=of i1] (i2);
				\node [right=1em of i2, blob] (b);
				\node [right=of b, blob] (c);
				\vertex [below=of f1] (f2);
				
				\diagram*{
				(i1) -- [fermion] (a) -- [fermion] (f1),
				(a) -- [scalar] (b),
				(a) -- [scalar] (c),
				};
				
			\end{feynman}
		
		\end{tikzpicture}
\caption{Feynman diagram corresponding to the insertion of a disformal vertex with any other arbitrary vertices}
\label{fig:disformal_general}
\end{figure}
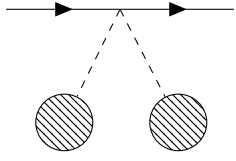
\begin{equation}
\mathrm{Fig} \ref{fig:disformal_general} = \int d\tau_A \left\langle T \left(\frac{d\phi}{d\tau_A} \right)^2 \mathcal{A} \mathcal{B} \right\rangle
\end{equation}
where $\mathcal{A}$ and $\mathcal{B}$ contain the other vertices. We can relate the proper time of each particle to the time of a distant observer \textit{via}
\begin{equation}
d\tau_\alpha = dt \sqrt{1-v_\alpha^2-h_{\mu \nu}v^\mu_\alpha v^\nu_\alpha}
\label{eq:proper_time}
\end{equation}
where $v^\mu_\alpha = \frac{dx_\alpha^\mu}{dt} = (1, \vec v_\alpha)$ and we have decomposed the metric as $g_{\mu \nu} = \eta_{\mu \nu} + h_{\mu \nu}$. There is a  map between $\tau_\alpha$ and $t$, whose precise form will depend on the trajectory of the two particles $\vec x_\alpha(t)$.

In the Feynman amplitude, one can contract the two fields coming from the two factors of  $d\phi/d\tau$ with two fields amongst the ones present in $\mathcal{A}$, $\mathcal{B}$, leading to the following expression
\begin{equation}
\int dt \; H(t) \frac{dF}{dt} \frac{dG}{dt}
\label{eq:general_disformal}
\end{equation}
where $F$, $G$ and $H$ are  functions of time depending on the vertices $\mathcal{A}$ and $\mathcal{B}$. These function can depend on time only through the trajectories of the two particles $\vec x_\alpha(t)$  but also on their velocities $\vec v_\alpha(t)$, accelerations $\vec a_\alpha(t)$ and possibly higher order derivatives. When computing the (conserved) two-body energy from the Lagrangian, one can replace the acceleration and their derivatives using the equations of motion which are of second order, thus retaining a dependence only through positions and velocities.

We will now show that the two functions $F$ and $G$ depend on time only through their relative position and velocities. In the case of a circular motion, these quantities depend very weakly on time, thus rendering the Feynman amplitude heavily suppressed. Let us now derive this statement.

At each order in perturbation theory one can integrate out the short distance gravitons $H_{\mu\nu}$ using the background field method, see \cite{Kuntz:2019zef, porto_effective_2016}, and keep the long wavelength ones $\bar h_{\mu\nu}$, where
$h_{\mu\nu}= H_{\mu\nu} +\bar h_{\mu\nu}$. Similarly one can integrate out the short distance modes of the scalar $\Phi$ and keep the long wavelength modes $\bar \phi$, see Sec. \ref{sec:multi}. The coupling of the long wavelength gravitons to matter defines the (pseudo) energy-momentum tensor ${\cal T}_{\mu\nu}$ (including the self-energy of the gravitational field) by
\be
S_{\rm grav}\supset \int d^4x \bar h_{\mu\nu} {\cal T}^{\mu\nu},
\ee
which is conserved at each order of perturbation theory thanks to diffeomorphism invariance applied to $\bar h_{\mu \nu}$.
The energy-momentum tensor allows us to define the centre-of-mass $\vec X$ of the system as
\begin{equation}
\left( \int d^3x \mathcal{T}^{00} \right) X^i = \int d^3x \mathcal T^{00} x^i.
\label{eq:center_of_mass}
\end{equation}
This comes from the invariance of the theory under boosts, which gives via Noether's theorem that the following charge is conserved
\begin{equation}
Q^{0i} = \int d^3x \left( \mathcal T^{00} x^i - \mathcal T^{0i} t \right).
\end{equation}
As the total momentum $P^i = \int d^3x \mathcal T^{0i}$ and energy $E = \int d^3x \mathcal T^{00}$ are also conserved, we get that the centre-of-mass moves with a constant velocity, thus justifying its definition. Even for GR and point-particle masses, the definition $\left( \sum_A m_A \right) X^i = \sum_A m_A x_A^i$ is valid only at lowest order in the PN expansion, since there are higher-order corrections implied by the formula \eqref{eq:center_of_mass}.

We can now use the two constraints on $\vec X$
\begin{equation}
\vec X = \frac{d \vec X}{dt} = \vec 0
\end{equation}
in order to relate the four unknows $(\vec x_\alpha, \; \vec v_\alpha)$ to the \textit{relative} coordinates $\vec x = \vec x_A - \vec x_B$ and $\vec v = \vec v_A - \vec v_B$. For example, in GR and in the case of the  circular motion of interest  the relation between coordinates and relative coordinates reads
\begin{equation}
m x_A^i = x^i \left[m_B + 3 \nu \delta m \left( \frac{G_NM}{r} \right)^2 \right] - \frac{4}{5} \frac{G_N^2 M^2 \nu \delta m}{r} v^i
\label{eq:CM_PN}
\end{equation}
and similarly for $\vec x_B$. Here we have introduced  the mass difference $\delta m = m_A - m_B$ and $r = |\vec x|$. This formula is valid up to 2.5PN order. The precise form of this relation is different for a scalar-tensor theory and has very recently been computed up to 3PN order in Ref \cite{Bernard:2018ivi}, but the statement that we can always recast the motion in terms of relative coordinates is not modified.

So finally, in the case of a circular motion, the two unknown functions $F$ and $G$ are functions of time through $\vec x$ and $\vec v$.
A scalar function built out of $\vec x$ and $\vec v$ must contain only $r$, $\vec x \cdot \vec v$ and $v$ by $SO(3)$ invariance. For circular motions,  $\dot{r} = \vec x \cdot \vec v = 0$ and also $\dot{v} = 0$ up to 2.5PN order\footnote{Actually, in a scalar-tensor theory, there could be a dipole energy loss term contributing at 1.5PN order to $\dot{r}$ \cite{Damour:1992we}. However, since we make the simplifying assumption of a universal scalar-tensor coupling $\beta$, such a dipole term vanishes.},
this means that the disformal contribution to the energy is indeed very suppressed.

We just showed that the time derivative of $\phi$ present in the disformal vertex \eqref{eq:disformal_vertex} behaves like
\begin{equation}
\frac{d\phi}{d\tau} \simeq \dot r \partial_r \phi + \dot v \partial_v \phi \simeq \frac{\dot r}{r} \phi + \frac{\dot  v}{v} \phi
\end{equation}
instead of the usual $\frac{d\phi}{d\tau} \sim \frac{v}{r} \phi$ that is expected from the post-Newtonian expansion (see App \ref{app:A}). From the energy balance between the Newtonian energy and gravitational wave emission one can find that in terms of  velocity power counting $\dot v = \mathcal{O} (v^6)$ and $\dot r= {\cal O}(v^6)$ too. If we then use the power-counting rules of App. \ref{app:A}, we find that the disformal vertex counts as
\begin{equation}
\frac{m}{\Lambda^2 m_\mathrm{Pl}^2} \int d\tau \left( \frac{d\phi}{d\tau} \right)^2 \sim \frac{v^{14}}{\Lambda^2 r^2}
\end{equation}
and the disformal diagram \ref{fig:disformal} counts as $L \beta^2 v^{14}/ (\Lambda^2 r^2)$ (where $L$ is the total angular momentum of the system, coming from the two vertices $\beta m/m_\mathrm{Pl} \int dt \phi$). This means that the disformal term is a 7PN effect in the conservative dynamics, so very highly suppressed.

% One can define both the effective mass of the system as
%\be
%{\cal M}= \int d^3 x\ {\cal T}_{00}
%\ee
%and the centre of mass position as
%\be
%{\cal M} X^i= \int d^3 x\ x^i  {\cal T}_{00}.
%\ee
%Conservation of the energy-momentum tensor $\partial_\mu {\cal T}^{\mu\nu}=0$  implies that $\frac{d{\cal M}}{dt}=0$ and $\dot P^i=0$ where $P^i= {\cal M} \frac{dX^i}{dt}$. As ${\cal T}_{\mu\nu}$ is built out of the positions and velocities of the two particles $A$ and $B$, the centre of mass and its velocities are functions of the positions $\vec x_\alpha$ and velocities $ \vec v_{\alpha}$. Imposing both that $\vec X=\vec 0$ and $ \vec P = \vec 0$ imposes two constraints in the mapping between the pairs $(\vec x_\alpha, \dot v _\alpha)$ and $(\vec x, \vec v, \vec X, \vec V)$ where $\vec P= {\cal M} \vec V$. As a result, one can express $\vec x_\alpha$ as a function of $\vec x$ and $\vec v$.

\section{Disformal radiation and Back-reaction}
\label{sec:disf}

\subsection{The multipole expansion of the dissipative dynamics}
\label{sec:multi}

Diagrammatic techniques allow us to find  the first disformal contribution to the radiated energy (into gravitational waves) of the system (we sketch in App \ref{app:B} how we could recover this result using directly the equations of motion). Let us focus on the scalar field  that is radiated away from the system, which we will call $\bar \phi$, i.e. it is a real scalar which is on shell. We will "integrate out" the conservative field which are involved in the interactions between the bodies and are not radiated away. We will denote this conservative field by $\Phi$, so that the total scalar field can be written as $\phi = \bar \phi + \Phi$. The explicit details of this procedure can be found in Ref \cite{Kuntz:2019zef}. The scalar action
becomes now
\begin{equation} \label{eq:scalar_coupling}
S_{\rm eff} \supset  \int d^4x  \left( -\frac{1}{2} \eta^{\mu \nu}  \partial_\mu \bar{\phi} \partial_\nu \bar{\phi}
+ \frac{1}{m_\mathrm{Pl}} J   \bar{\phi} \right)
\end{equation}
where we have neglected interacting terms quadratic or higher in the radiated fields that contribute only at high order in the velocity expansion \cite{goldberger_gravitational_2010}. We will give the explicit expression of the scalar coupling $J$ in the following, but before doing this we will relate it to the emitted power by using a multipole expansion. As we are interested in physical configurations where the radiated scalar field $\bar \phi$ varies on scales much larger than the size of the  source, we can expand the scalar field around the centre-of-mass of the source (i.e. we set $\vec X = \vec 0$) up to second order
\begin{equation}
\bar \phi (t, \vec x) = \bar{\phi}(t, \vec 0) + x^i \partial_i \bar{\phi}(t,\vec 0) + \frac{1}{2} x^i x^j \partial_i \partial_j \bar{\phi}(t, \vec 0) \; .
\end{equation}
As the radiated field satisfies the scaling $\partial_i \bar \phi \sim \frac{v}{r} \bar \phi$, this is nothing but a velocity expansion. Then by rearranging terms in the scalar coupling \eqref{eq:scalar_coupling} (using only symmetric traceless multipoles), we can write the interacting term up to second order in the velocity expansion as
\begin{equation}
S_{\rm int}^{(\phi)} =  \frac{1}{m_\mathrm{Pl}} \int dt \left( I_\phi \bar{\phi} + I_\phi^i \partial_i \bar{\phi} + \frac{1}{2} I_\phi^{ij} \partial_i \partial_j \bar{\phi}  \right)
\end{equation}
where the multipole moments are given by
\begin{equation} \label{eq:scalar_dipole_quadrupole}
I_\phi  \equiv  \int d^3 x \left( J + \frac{1}{6}  \partial_t^2 J  x^2 \right) \;, \quad I_\phi^i  \equiv \int d^3x J x^i \;, \quad  I_\phi^{ij} \equiv \int d^3x J \left(  x^i x^j - \frac{1}{3}x^2 \delta^{ij}\right) \; .
\end{equation}
From this expression, the power radiated into the scalar field can be found
\begin{equation} \label{eq:radiated_power_scalar}
P_\phi = {2G_N}\left[  \big\langle \dot{I}_\phi^2  \big\rangle  + \frac{1}{3} \big\langle  \ddot{I}_\phi^i \ddot{I}_\phi^i  \big\rangle + \frac{1}{30} \big\langle \dddot{I}_\phi^{ij}  \dddot{I}_\phi^{ij}   \big\rangle \right]
\end{equation}
where the brackets denote the average over many gravitational wave cycles. This is the scalar counterpart of the GR power radiated into gravitons
\begin{equation} \label{eq:quadrupole}
P_h = \frac{G_N}{5}  \left\langle \dddot{I}_h^{ij}\  \dddot{I}_h^{ij} \right\rangle
\end{equation}
where $I_h^{ij} \equiv \int d^3x T^{00} \left( x^i x^j - \frac{1}{3}x^2 \delta^{ij}\right)$ is the gravitational quadrupole of the source.
These formulae are derived in Ref \cite{Kuntz:2019zef}.

 To lowest order in the velocity expansion, $J$ is simply given from eq. \eqref{eq:disformal_vertex} by
\begin{equation} \label{eq:J_v0}
J_{v^0} = - \beta \left( m_A \delta^3(\vec{x}-\vec{x}_A) + m_B \delta^3(\vec{x}-\vec{x}_B) \right)
\end{equation}
where $A$ and $B$ denote the two objects that we are referring to.
The first relativistic correction to the source $J$ comes by integrating out the conservative fields. Following \cite{Kuntz:2019zef}, the conformal part of the coupling at second order is given by
\begin{equation}
J_{v^2} = \beta \left( m_A \frac{v_A^2}{2} \delta^3(\vec{x}-\vec{x}_A) + (A \leftrightarrow B) \right) + \beta \frac{G_Nm_Am_B}{|\vec x_A - \vec x_B|} \left( \delta^3(\vec{x}-\vec{x}_A) + (A \leftrightarrow B) \right) \; .
\end{equation}
The first disformal contribution to $J$ involves only one Feynman diagram, Figure \ref{fig:J_v2}. The detailed calculation of this diagram are in App \ref{app:A}. This  gives the following contribution to $J$
\begin{figure}
	\centering
		\begin{tikzpicture}
			\begin{feynman}
				\vertex (i1);
				\vertex [right=of i1] (a);
				\vertex [right=of a] (f1);
				\vertex [above=of f1] (s);
				\vertex [below=of i1] (i2);
				\vertex [below=of a] (b);
				\vertex [below=of f1] (f2);
				
				\diagram*{
				i1 -- [fermion] (a) -- [fermion] (f1),
				(a) -- [gluon] (s),
				(i2) -- [fermion] (b) -- [fermion] (f2),
				(a) -- [scalar] (b)
				};
			\end{feynman}
		
		\end{tikzpicture}

\caption{Feynman diagram contributing to the emission of one radiation scalar, at  order $v^2$. It should be added to its symmetric counterpart. The curly line represents the radiated field $\bar \phi$ while the dotted line represents the potential field $\Phi$. The vertex shown in this diagram is written in eq. \eqref{eq:dissipative_vertex}}
\label{fig:J_v2}
\end{figure}
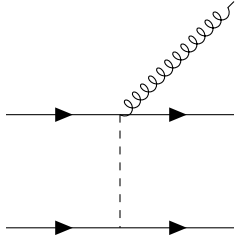
\begin{equation}
J^\mathrm{disf} = 4 \beta \frac{G_Nm_Am_B}{\Lambda^2} \frac{d^2}{dt^2} \frac{1}{|\vec x_A - \vec x_B|} \left( \delta^3(\vec{x}-\vec{x}_A) + (A \leftrightarrow B) \right) \; .
\end{equation}
Notice that, as for the conservative part, this contribution vanishes exactly for a circular orbit.
The lowest-order contribution of the disformal source term $J^\mathrm{disf}$ in the multipole expansion is in the monopole (higher multipoles are further velocity-suppressed). We can consequently write the dipole and quadrupole emission terms using only the expression for $J_{v^0}$
\begin{align}
\begin{split}
I_\phi^{ij} &= - \beta \left( m_A \left( x^i x^j - \frac{1}{3}x^2 \delta^{ij}\right) + (A \leftrightarrow B) \right) \\
I_\phi^{i} &= - \beta \left( m_A x_A^i + (A \leftrightarrow B) \right)
\end{split}
\end{align}
and their contribution in the radiated power will be the same as in conformally coupled scalar-tensor theories. As we focus on a universal conformal coupling $\beta$, the contribution of the dipole in the radiated power vanishes, because its second derivative is zero by the centre-of-mass theorem at lowest order in the velocity. The lowest order contribution to the monopole, given by $J_{v^0}$, vanishes also because it is simply $I_{\phi, \; v^0} = - \beta (m_A + m_B)$ which is a constant by conservation of matter. The next contribution to the monopole, including the disformal contribution, starts at the same order as the quadrupole radiation in the velocity expansion and can be written as
\begin{align}
\begin{split}
I_{\phi, \; v^2 + \mathrm{disf}} &= \frac{\beta}{6} (m_A v_A^2 + m_B v_B^2) + \frac{\beta}{3}(7+2\beta^2) \frac{G_Nm_1m_2}{|\vec x_A - \vec x_B|} \\
&+ 8 \beta \frac{G_Nm_Am_B}{\Lambda^2} \frac{d^2}{dt^2} \frac{1}{|\vec x_A - \vec x_B|} \; .
\end{split}
\end{align}
At this order  $m_A \frac{v_A^2}{2} + m_B \frac{v_B^2}{2} \simeq \frac{G_N(1+2\beta^2)m_Am_B}{|\vec x_A - \vec x_B|}$ by conservation of the Newtonian energy hence one can rewrite the monopole term as
\begin{equation} \label{eq:monopole}
I_{\phi, \; v^2 + \mathrm{disf}} = 4 \beta G_N m_Am_B \left( \frac{2+\beta^2}{3|\vec x_A - \vec x_B|} + \frac{2}{\Lambda^2} \frac{d^2}{dt^2} \frac{1}{|\vec x_A - \vec x_B|} \right).
\end{equation}
At this point one can easily see that the disformal term is a $v^2/(\Lambda^2 r^2)$ correction to the conformal monopole, i.e a 1PN effect for an elliptic orbit. In the case of a circular orbit, we showed in Sec. \ref{sec:circular} that one should replace $d/dt \rightarrow v^6/r$ instead of the usual counting $d/dt \rightarrow v/r$, and so the disformal term is a $v^{12}/(\Lambda^2 r^2)$, i.e 6PN, correction to the conformal monopole.

In the next Section we will then use the relation $P_\phi = 2 G_N \left \langle \dot I_\phi^2 \right \rangle$ to find the final expression for the radiated power.

\subsection{Radiated power of elliptic orbits}

In this Section we will calculate the power emitted from the system in an eccentric orbit.
As discussed before, the total emitted power  splits into the power lost into gravitons and the one lost into the scalar field. For the graviton case the power emitted for elliptic orbits is known from the Peter-Mathews formula\cite{PhysRev.131.435}. Here one should use the normalised Newton constant $\tilde G$ instead of $G_N$ in the quadrupole moment of the source. Using eq. \eqref{eq:quadrupole} we obtain for this power
\begin{align}
\begin{split}
P_h &= \frac{32}{5\tilde{G} (1+2\beta^2)} (\tilde{G} M_c \omega)^{10/3} f_1(e) \;, \\
f_1(e) &= \frac{1}{(1-e^2)^{7/2}}(1+\frac{73}{24}e^2+\frac{37}{96}e^4)
\end{split}
\end{align}
where $M_c = (m_Am_B)^{3/5}/M^{1/5}$ is the chirp mass, $M=m_A + m_B$ is the total mass of the system, and $\omega = \frac{2\pi}{T}$ is the frequency of the system that satisfies Kepler's third law
\begin{equation}\label{eq:Kepler}
\omega^2 = \frac{\tilde G M}{a^3} \; .
\end{equation}

Let us now discuss the scalar radiation. Since the scalar quadrupole \eqref{eq:scalar_dipole_quadrupole} is proportional to  the gravitational quadrupole, we deduce from eq. \eqref{eq:radiated_power_scalar} that the scalar quadrupole power loss is
\begin{equation}
P_\phi^\mathrm{quad} = \frac{\beta^2}{3} P_h\; .
\end{equation}
As discussed above, the scalar dipole is zero as we focus on a universal scalar coupling $\beta$. We are left  to calculate the monopole power starting from eq. \eqref{eq:monopole}. Using the same method as in Sec. \ref{sec:planar_dynamics}, we can define the average over many gravitational wave cycles appearing in the emitted monopole power \eqref{eq:radiated_power_scalar} as
\begin{align}
\begin{split}
\left \langle \dot I_\phi^2 \right \rangle = \frac{1}{2\pi} \int_0^{2\pi} d\psi \frac{(1-e^2)^{3/2}}{(1+e\cos \psi)^2} \dot I_\phi^2(\psi)
\end{split}
\end{align}
where $T$ is the period of the system, $e$ the excentricity and $\psi$ is the angle along the trajectory defined in eq. \eqref{eq:elliptic_trajectory}. This integral leads to the monopole power
\begin{align}
\begin{split}
P_\phi^\mathrm{mono} &= \frac{16}{9\tilde G} \frac{\beta^2(2+\beta^2)^2}{(1+2\beta^2)^3} (\tilde G M_c \omega)^{10/3} \\
&\times \left(f_2(e)-12yf_3(e)+36y^2f_4(e) \right)
\end{split}
\end{align}
where the three ellipticity functions $f_2$, $f_3$ and $f_4$ are given by :
\begin{align}
\begin{split}
f_2(e) &= \frac{e^2}{(1-e^2)^{7/2}} \left(1+\frac{1}{4} e^2\right) \\
f_3(e) &= \frac{e^2}{(1-e^2)^{13/2}} \left(1+\frac{37}{4}e^2+\frac{59}{8}e^4 + \frac{27}{64}e^6 \right) \\
f_4(e) &= \frac{e^2}{(1-e^2)^{19/2}} \left(1+\frac{217}{4}e^2+\frac{1259}{4}e^4+\frac{11815}{32}e^6+\frac{11455}{128}e^8+\frac{1125}{512}e^{10} \right)
\end{split}
\end{align}
and $y$ is the parameter
\begin{equation} \label{eq:def_y}
y = \frac{\tilde G M}{(2+\beta^2)\Lambda^2a^3} = \frac{1}{2+\beta^2} \left(\frac{\omega}{\Lambda} \right)^2
\end{equation}
where for the second equality we have used Kepler's third law \eqref{eq:Kepler}.

The first ellipticity function corresponds to the conformal case without any disformal interaction and agrees with other references \cite{damour_tensor-multi-scalar_1992}, whereas the other functions contribute only when a disformal interaction is present. We can remark that, in order for the disformal correction to be small, we have to impose $y\lesssim 1$, i.e $\omega \lesssim \Lambda$. This is a reformulation of the scaling that we obtained in the App. \ref{app:A} by examining the weight of the disformal operator in the PN expansion. One can also relate the parameter $y$ to the nonlinear radius introduced in eq. \eqref{eq:nonlinear_radius} via
\begin{equation}
y \simeq \frac{1}{\beta^2} \frac{r_\star}{r_s} \left( \frac{r_*}{a} \right)^3
\end{equation}
in terms of both the nonlinear radius $r_*$ and the Schwarzschild radius $r_s = 2G_N M$ associated to the total mass $M$. Notice that as long as $a\gtrsim r_\star$, this should be smaller than unity. At the limit of validity of the perturbative treatment when $a\simeq r_\star$, this can become of order unity when the non-linear radius is $r_\star \gtrsim \beta^2 r_s $, i.e. when the disformal scale $\Lambda$ is not too large.

By summing up all contributions, we find the final formula for the emitted power :
\begin{align} \label{eq:disformal_power_total}
\begin{split}
P &= \frac{32}{5 (1+2\beta^2) \tilde G} (\tilde G M_c \omega)^{10/3} \\
& \times \left( (1+\frac{\beta^2}{3})f_1(e)+\frac{5}{18} \frac{\beta^2(2+\beta^2)^2}{(1+2\beta^2)^2}\left(f_2(e)-12yf_3(e)+36y^2f_4(e) \right) \right) \; .
\end{split}
\end{align}
Notice that the disformal contribution vanishes for circular orbits. As for the conservative dynamics, the leading contribution in this case is much suppressed compared to the monopole and quadrupole due to the conformal interaction.

\subsection{Constraint from the Hulse-Taylor pulsar} \label{sec:Hulse_Taylor}

As stated above, the observation of an elliptic inspiral in a GW detector could allow us to put constraints on a disformal interaction. The correction to the total energy \eqref{eq:disformal_energy_elliptic} and to the dissipated power \eqref{eq:disformal_power_total} should be consistently used to derive a waveform template in order to perform a matched filter analysis. We can note from eq. \eqref{eq:def_y} that the strongest constraint would come from systems with high frequencies, i.e from elliptic systems in the LIGO/Virgo band. While the majority of such systems are expected to have circular orbits, it could also be possible to observe an eccentric merger induced by, e.g., Kozai oscillations of a triple system \cite{Kozai:1962zz, Wen:2002km}.

In this Section we will rather focus on the simple constraint coming from the observation of the Hulse-Taylor pulsar B1913+16, which is well-known to have an orbital decay consistent with GR at the 0.2 percent level \cite{Hulse:1974eb, Weisberg:2004hi}. This system does also have a  large eccentricity, $e \simeq 0.6$, which does maximise the disformal effect. We use the parameters inferred from the non-relativistic analysis of arrival time data quoted in Ref. \cite{Weisberg:2004hi} (we only need the orbital period $P_b = 2\pi /\omega$ and the eccentricity $e$). By simply requiring that the dissipated power in eq. \eqref{eq:disformal_power_total} should not be corrected by more than $0.2$ percent by the disformal effect, we find the following bound on $\Lambda$
\begin{equation}
\Lambda \gtrsim 10^{-17} \; \mathrm{eV}
\end{equation}
where we have also used the fact that the conformal coupling is strongly constrained by the Cassini bound, $\beta^2 \lesssim 10^{-5}$. This bound is comparable with the one from torsion pendulum experiments \cite{Brax:2014d}.

\section{Conclusion} \label{sec:conclusion}

We have analysed the first correction to the conservative and dissipative parts of the effective action for a two body system coupled both conformally and disformally via a massless scalar field. For circular orbits, the effect of the disformal coupling is very suppressed in the PN expansion. For elliptical orbits, the corrections are small in the perturbative regime corresponding to large separation between the objects in the inspiralling phase. When the two bodies approach each other, the effects can become of order unity when perturbation theory starts breaking down. Hence the disformal interaction playing a significant role in the dynamics of the two body system at the end of the inspiralling phase and during the merging cannot be excluded. Of course this is beyond the present scope of this paper, and should be investigated as more data will eventually allow one to test extensions of General Relativity
more deeply.

\acknowledgments
This work is
supported in part by the EU Horizon 2020 research and innovation
programme under the Marie-Sklodowska grant No. 690575. This article is
based upon work related to the COST Action CA15117 (CANTATA) supported
by COST (European Cooperation in Science and Technology). ACD acknowledges partial support from STFC under grants ST/L000385 and ST/L000636.
 It is a pleasure to thank Federico Piazza and Filippo Vernizzi for useful discussions. We are grateful to F\'elix Juli\'e for comments and for noticing a few typographical mistakes in our manuscript.

\appendix

\section{Diagrammatic expansion\label{app:A}}

\subsection{Feynman rules}

Feynman diagrams allow one to derive the effective action  by integrating out the scalar and gravitational fields
\begin{equation}
\exp\left({i S_{\rm eff}[x_\alpha]} \right) = \int {\cal D} h_{\mu \nu} {\cal D} \phi \exp \left({i  S[x_a, {h}_{\mu \nu}, {\phi}] + i S_{{\rm GF}, h} [h_{\mu \nu}] } \right)
\label{eq:path_integral}
\end{equation}
Here we have split the metric according to $g_{\mu \nu} = \eta_{\mu \nu} + h_{\mu \nu}$ and $S_{{\rm GF}, h}$ refers to a gauge-fixing term that is needed in order to make the graviton propagator well-defined. As we will only consider diagrams with scalars here, we will not need its expression. The path-integral generates both a real and an imaginary part in the effective action. The former corresponds to the conservative dynamics of the objects and the latter is related to the energy loss of the system by the optical theorem \cite{porto_effective_2016}. We will concentrate here on the conservative dynamics  first. The effective action is associated with the action of two objects of trajectory $x_\alpha(t)$
\begin{equation}
S_{\mathrm{eff}} = \int dt L[\mathbf{v}_\alpha, \mathbf{x}_\alpha].
\end{equation}
In order to compute $S_\mathrm{eff}$ as an expansion in terms of Feynman diagrams, we need the scalar propagator coming from the quadratic term in the action $-\int d^4x \sqrt{-g} (\partial \phi)^2$
\begin{equation}
\langle T \phi_{\vec{k}} (t) \phi_{\vec q} (t') \rangle = - (2 \pi)^3 \frac{i}{k^2} \delta^{(3)}(\vec k + \vec q) \delta(t-t')
\label{eq:propagator}
\end{equation}
where we have imposed that the spatial part of the momentum $\vec{k}$ is much greater than the temporal one $k^0$. This can be understood as the fact that velocity-dependent terms come only as a perturbation of the static potential in a non-relativistic treatment.
The Feynman rules generated by the path-integral \eqref{eq:path_integral} (taking into account only potential fields, i.e that do not radiate energy outside of the system) are the following :
\begin{itemize}

\item Draw all the diagrams that remain connected when removing the world-lines of the particles, discarding quantum loop diagrams.

\item Divide by the symmetry factor of the diagram

\item For each vertex (the relevant ones in this article being given in eq. \eqref{eq:disformal_vertex}), multiply the corresponding expression by $i$.

\item Contract all the internal gravitons. This gives a combinatorial factor corresponding to the number of Wick contractions, plus the propagator given in eq. \eqref{eq:propagator}.

\end{itemize}
Let us mention the power-counting rules associated with the vertices. The expansion parameter of Non-Relativistic General Relativity is, as in a Post-Newtonian expansion, the typical velocity of the two bodies $v$ which can be related to the orbital radius $r$ \textit{via} the virial theorem $v^2 \sim G_Nm/r$. Each vertex in the action \eqref{eq:disformal_vertex} can be associated to a weight in this expansion \cite{goldberger_effective_2006}. The counting goes as follows :
\begin{itemize}
\item A scalar or graviton counts as $\sqrt{v}/r$, as can be obtained from the propagator.
\item The virial theorem states that $m_\alpha /m_\mathrm{Pl}$ counts as $\sqrt{Lv}$ where $L = mrv$ is the angular momentum of the system.
\item Times counts as $t \sim r/v$.
\item Finally, each diagram should be proportional to $L$, as it is the loop counting parameter of the system and we consider only classical dynamics.
\end{itemize}
Using these rules, we can see that the lowest-order scalar vertex counts as $m/m_\mathrm{Pl} \int dt \phi \sim \sqrt{L}$ which gives the Newtonian diagram
\begin{equation}
\frac{m^2}{m_\mathrm{Pl}^2} \int dt_1 dt_2 \left\langle T \phi \phi \right\rangle \sim L
\end{equation}
while the disformal vertex counts as
\begin{equation}
\frac{m_\alpha}{\Lambda^2 m_\mathrm{Pl}^2} \int d\tau_\alpha \left( \partial_\mu \phi u^\mu_\alpha \right)^2 \sim \frac{v^4}{\Lambda^2 r^2}
\end{equation}
so that, if we impose that it should be of the same order than a typical term entering the first Post-Newtonian corrections of the energy like $m/m_\mathrm{Pl}^2 \int d\tau \phi^2$ (which counts as $v^2$) in order to have an interesting perturbative deviation from GR,  we have the following bound on $\Lambda$ :
\begin{equation} \label{eq:bound_Lambda}
r\gtrsim \frac{v}{\Lambda}
\end{equation}

The precise bound on $\Lambda$ coming from observations using this estimate is given in Sec. \ref{sec:Hulse_Taylor}, where we examine the constraint coming from the disformal radiation in the Hulse-Taylor binary pulsar.

%which is a stringent bound for $v\sim 10^{-3}$ and $\Lambda \gtrsim 10^{-17}$ eV corresponding to  $r\gtrsim 10 000$ km. For the Hulse-Taylor binary pulsar where the size of the orbit is around $600 000$ km, the approximation is valid.

\subsection{First conservative correction}

We now calculate the diagram given in Figure \ref{fig:disformal} using these rules. Using the expression of the proper time given in eq. \eqref{eq:proper_time}  we find :
\begin{equation}
\mathrm{Fig} \ref{fig:disformal} = \frac{1}{2} \frac{im_1}{\Lambda^2 m_\mathrm{Pl}^2} \int dt_1 \frac{(-i)^2\beta^2m_2^2}{m_\mathrm{Pl}^2} \int dt_2 dt_3 \left\langle T (\partial_\mu \phi(t_1, x_1) u_1^\mu)^2 \phi(t_2, x_2) \phi(t_3, x_3) \right\rangle
\end{equation}
where the $1/2$ is the symmetry factor of the diagram, and for simplicity we denote by $1$ and $2$ the two bodies (instead of $A$, $B$ in the rest of the article).
Using the Fourier transform of the field and Wick's contraction we have
\begin{align}
\begin{split}
\mathrm{Fig} \ref{fig:disformal} &= i \frac{\beta^2m_1m_2^2}{\Lambda^2m_\mathrm{Pl}^4} \int dt_1 dt_2 dt_3 \frac{d^3k_1 d^3k_2}{(2\pi)^6} \frac{1}{k_1^2k_2^2} e^{i\vec{k}_1 \cdot (\vec x_1(t_1) - \vec x_2(t_2))} e^{i\vec{k}_2 \cdot (\vec x_1(t_1) - \vec x_2(t_3))} \\
& \times \left[\frac{d}{dt_1}\delta(t_1-t_2) + \delta(t_1-t_2)i\vec k_1 \cdot \vec v_1 \right] \left[\frac{d}{dt_1}\delta(t_1-t_3) + \delta(t_1-t_3)i\vec k_2 \cdot \vec v_1 \right]
\end{split}
\end{align}
Using $\frac{d}{dt_1}\delta(t_1-t_2) = - \frac{d}{dt_2}\delta(t_1-t_2)$ and integrating by parts, we have
\begin{align}
\begin{split}
\mathrm{Fig} \ref{fig:disformal} &= -i \frac{\beta^2m_1m_2^2}{\Lambda ^2m_\mathrm{Pl}^4} \int dt \frac{d^3k_1 d^3k_2}{(2\pi)^6} \frac{1}{k_1^2k_2^2} e^{i\vec{k}_1 \cdot (\vec x_1 - \vec x_2)} e^{i\vec{k}_2 \cdot (\vec x_1 - \vec x_2)} \\
& \times \left[\vec k_1 \cdot (\vec v_1 - \vec v_2) \right] \left[\vec k_2 \cdot (\vec v_1 - \vec v_2) \right]
\end{split}
\end{align}
Finally with the formula $\int \frac{d^3k}{(2\pi)^3} \frac{\vec k}{k^2} e^{i \vec k \cdot \vec r} = i \frac{\vec r}{4\pi r^3}$, with $\vec r = \vec x_1 - \vec x_2$ we get
\begin{equation}
\mathrm{Fig} \ref{fig:disformal} = i \frac{\beta^2m_1m_2^2}{16 \pi^2 \Lambda^2m_\mathrm{Pl}^4} \int dt \frac{(\vec n \cdot (\vec v_1 - \vec v_2))^2}{r^4}
\end{equation}
with $\vec n = \vec r/r$. Upon using Planck's mass definition $G_N = \frac{1}{8\pi m_\mathrm{Pl}^2}$ and summing over the symmetric diagram, we finally find the formula for the first disformal correction to the two-body Lagrangian given in eq. \eqref{eq:disformal_energy}.

\subsection{First dissipative correction}

The second diagram that we have to calculate is shown in Figure \ref{fig:J_v2}, and gives the contribution of the disformal vertex to the dissipative dynamics. Decomposing the total field into a conservative and a dissipative part as $\phi = \bar \phi + \Phi$, this diagram emerges from the vertex coupling $\bar \phi$ and $\Phi$, which can be written (using eq. \eqref{eq:disformal_vertex}) as :
\begin{equation} \label{eq:dissipative_vertex}
2 \frac{m_\alpha}{\Lambda^2 m_\mathrm{Pl}^2} \int dt \partial_\mu \bar \phi u_\alpha^\mu  \partial_\mu \Phi u_\alpha^\mu
\end{equation}
Here $\bar \phi$ is treated as an external field while $\Phi$ is integrated out and enters the Feynman diagrams in the internal lines. Using this expression, the diagram of Figure \ref{fig:J_v2} can be written as
\begin{equation}
\mathrm{Fig} \ref{fig:J_v2} = 2i\frac{m_1}{\Lambda^2 m_\mathrm{Pl}^2} \int dt_1 (-i) \beta \frac{m_2}{m_\mathrm{Pl}} \int dt_2 \partial_\mu \bar \phi(t_1, \vec x_1) u_1^\mu \left \langle T \partial_\mu \Phi(t_1, \vec x_1) u_1^\mu \Phi(t_2, \vec x_2) \right \rangle
\end{equation}
Using manipulations similar to the calculation of the previous diagram, one finds that this gives
\begin{equation}
\mathrm{Fig} \ref{fig:J_v2} = 2i \beta \frac{m_1m_2}{\Lambda^2 m_\mathrm{Pl}^3} \int dt \partial_\mu \bar \phi(t_1, \vec x_1) u_1^\mu \frac{\vec n \cdot (\vec v_1 - \vec v_2)}{4\pi r^2}
\end{equation}
To this expression one should add the symmetric diagram obtained by exchanging labels 1 and 2. By noticing that $\partial_\mu \bar \phi(t_1, \vec x_1) u_1^\mu = \frac{d}{dt} \bar \phi(x_1)$, one can integrate by parts in order to put this diagram in a form similar to $i\int d^4x \frac{J}{m_\mathrm{Pl}} \bar \phi$ as required by the equation \eqref{eq:scalar_coupling} defining $J$. By further noticing that $\frac{\vec n \cdot (\vec v_1 - \vec v_2)}{r^2} = - \frac{d}{dt} \frac{1}{r}$, one can write the formula for the disformal contribution to $J$ that is found in the main text
\begin{equation}
J^\mathrm{disf} = 4 \beta \frac{G_Nm_Am_B}{\Lambda^2} \frac{d^2}{dt^2} \frac{1}{|\vec x_A - \vec x_B|} \left( \delta^3(\vec{x}-\vec{x}_A) + (A \leftrightarrow B) \right)
\end{equation}

\section{Radiation from point sources} \label{app:B}

Energy loss implies that  bound systems on elliptical orbits inspiral towards each other.
The power radiated by the sources is given by
\be
P=\int d^2 S^i T_{0i}^\phi
\ee
across a surface far away from the point sources. This is the flux due to the scalar energy momentum tensor.  Using the divergence theorem this is nothing but
\be
P= \int d^3 x \partial^i T_{0i}^\phi
\ee
Using the conservation of the total energy momentum tensor, this can be rewritten as
\be
P= -\int d^3 x \ \partial^\mu T_{0\mu} - \frac{d}{dt} \int d^3 x T_{00}^\phi.
\ee
The Bianchi identity and the Einstein equation imply the non-conservation equation for the scalar energy-momentum tensor
\be
\partial^\mu T_{\mu\nu}= \frac{\beta T}{m_{\rm Pl}} \partial_\nu \phi - \frac{2}{M^4} \partial_\mu(\partial_\lambda\phi  T^{\mu\lambda}) \partial_\nu \phi.
\ee
Focussing on a two body system where orbits are closed, and taking the average  over the orbit, the last term of the previous equation averages to zero implying that
\be
\left \langle P\right \rangle =- \frac{\beta}{m_{\rm Pl}} \int d^3 x \left\langle T \frac{d\phi}{dt}\right\rangle + \frac{2}{M^4} \int d^3 x \left \langle\partial_\mu ( \partial_\lambda\phi  T^{\mu\lambda}) \frac{d \phi}{dt}\right\rangle
\ee
We have  $\phi= \phi^{(0)} +\delta \phi^{(0)}$ where
\be
\phi^{(0)}= -\Box^{-1}_{\rm ret}( \frac{\beta T}{m_{\rm Pl}})=\frac{\beta}{4\pi m_{\rm Pl}} \int d^3 x' \frac{1}{\vert  \vec x -\vec x'\vert} T(\vec x',x^0 -\frac{\vert \vec x -\vec x'\vert}{c})
\ee
and  the inverse d'Alembertian is calculated using the retarded Green's function $G(x,x')= -\frac{1}{2} \theta (x^0- x^{'0}) \delta ((x-x')^2)$
where  $x^2= -c^2 t^2 +\vec x^2$. Notice we have reinstated the speed of light $c$ as a book-keeping parameter. Similarly we have
\be
\delta \phi^{(0)}= \frac{2}{m_{\rm Pl}^2 \Lambda^2} \Box^{-1}_{\rm ret}  ( \partial_\mu (  \partial_\nu\phi^{(0)} T^{\mu\nu}))=-\frac{1}{2\pi m_{\rm Pl}^2 \Lambda^2}
\int d^3 x' \frac{1}{\vert  \vec x -\vec x'\vert} (\partial_\mu (  \partial_\nu\phi^{(0)} T^{\mu\nu}))(\vec x',x^0 -\frac{\vert \vec x -\vec x'\vert}{c}).
\ee
The dissipated power contains three terms. The first one comes from the conformal coupling only and reads
\be
\left \langle P\right \rangle_{\rm conf} =- \frac{\beta}{m_{\rm Pl}} \int d^3 x \left\langle T \frac{d\phi^{(0)}}{dt}\right\rangle
\ee
which after integrations by parts becomes
\be
\left \langle P\right \rangle_{\rm conf}=\frac{\beta^2 }{4\pi m_{\rm Pl}^2}\sum_{n\ge 1} \frac{(-1)^{n+1}}{c^{2n+1}} \int d^3 x d^3 x' \frac{d^{n+1} }{dx_0^{n+1}} T(\vec x,x^0) \vert \vec x- \vec x'\vert^{2n} \frac{d^{n+1} }{dx_0^{n+1}} T(\vec x',x^0).
\ee
All the terms involving even inverse powers of $c$ vanish in this series. The term corresponding to $n=0$ in $c^{-1}$ vanishes too as mass is conserved.
The disformal contributions to the emitted power are new and read
\be
\left \langle P\right \rangle_{\rm dif}=- \frac{\beta}{m_{\rm Pl}} \int d^3 x \left\langle T \frac{d\delta\phi^{0}}{dt}\right\rangle + \frac{2}{m_{\rm Pl^2}\Lambda^2} \int d^3 x \left \langle\partial_\mu ( \partial_\lambda\phi^{(0)}  T^{\mu\lambda}) \frac{d \phi^{(0)}}{dt}\right\rangle.
\ee
The first term involves the correction to $\phi$ coming from the disformal interaction whilst the second is the direct effect of the disformal interaction in the conservation of
energy-momentum. In the main text, the conformally and disformally emitted powers have  been evaluated using diagrammatic techniques.

\bibliographystyle{JHEP}
\bibliography{bib}

\end{document}